\documentstyle[aps, aps12]{revtex}
\begin{document}

\title{Severe Limits on Variations of the Speed of Light with Frequency}
\draft
\author{Bradley E. Schaefer}
\address{Physics Department, Yale University, PO Box 208121, New Haven CT 
06520-8121}
\address{schaefer@grb2.physics.yale.edu}
\date{\today}
\maketitle
\begin{abstract}

Explosive astrophysical events at high red shift can be used to place severe 
limits on the fractional variation in the speed of light ($\Delta c/c$),
the photon mass ($m_{\gamma}$), and the energy scale of quantum gravity
($E_{QG}$). I find $\Delta c/c < 6.3 \times 10^{-21}$ based on the
simultaneous arrival of a
flare in GRB 930229 with a rise time of $220 \pm 30 \mu s$ for photons of
30 keV and 200 keV. The limit on $m_{\gamma}$ is $4.2 \times 10^{-44} g$
for GRB 980703 from radio to gamma ray observations.  The limit on
$E_{QG}$ is $8.3 \times 10^{16}$ GeV for GRB 930131 from 30 keV to 80 MeV
photons.

\end{abstract}

The question of whether the speed of light varies with frequency is of 
fundamental and current interest:  1) Einstein's postulate of the
invariance of light is the cornerstone of much of modern physics, so tests
of the correctness of this postulate should be pushed as far as possible.
2) With the recent evidence from Super-Kamiokande that the neutrino likely
has a mass \cite{fukuda}, the question of the mass of the photon should be
re-examined.  3) Quantum gravity models suggest \cite{camelia}
\cite{gambini} that the speed of light has an effective energy dependence.

Classical textbooks \cite{jackson} and review articles \cite{hall} report
the status as of the middle 1970's.  Laboratory and accelerator
experiments were not able test for fractional variations in the speed of
light ($\Delta c/c$) to better than roughly $10^{-8}$. The first limit
that took advantage of astronomical distances compared the arrival time of
radio and optical emission from flare stars to constrain $\Delta c/c < 
10^{-6}$ \cite{lovell}. Then B. Warner and R. Nather measured the phase
difference for pulses of the Crab Pulsar between optical wavelengths of
0.35 and 0.55 microns to be less than 10 microseconds \cite{warner}.  At a
distance of 2 kpc, this limits $\Delta c/c$ to be less than $5
\times 10^{-17}$. 

Surprisingly, no further improvements on $\Delta c/c$ have been made on
the Crab Pulsar limit, and the topic has received little subsequent
discussion in the literature. Indeed, the speed of light has been defined
to be a constant for purposes of metrology. In the meantime, a key
assumption for the Crab Pulsar limit has been severely undermined since
five out of six pulsars detected at high energy have pulse structures that
vary strongly both in shape and phase as a function of frequency
\cite{thompson}. Within the last year, several groups have independently
realized that Gamma Ray Bursts (GRBs) provide a means to look for delays
in light traversing extremely large distances.  Amelino-Camelia {\em et
al.} \cite{camelia} set approximate limits on the dispersion scale for
quantum gravity, while Biller {\em et al.} \cite{biller} set stricter
limits by analysis
of a short TeV flare seen in a nearby active galaxy (Mkn 421). 

In this paper, I present new limits based on a variety of explosive events 
at high red shifts.  My original motivation was the discovery last year of
a flare in a Gamma Ray Burst (GRB 930229) with a $220 \pm 30 \mu s$
rise time that occurs simultaneously from 30 to 200 keV \cite{schaefer},
and the realization that this constrains the dispersion of light to be
less than a millisecond out of a Hubble Time.  However, the relevant limit
depends on the assumed functional form for the frequency dependence of
`c', so different events are the most restrictive in the various cases.

In Table 1, I have gathered the data for the most restrictive events of
various classes.  These include short duration GRBs, GRBs with GeV
photons, GRBs with associated x-ray/optical/radio transients, high red
shift Type Ia supernovae, the active galaxy Mkn 421, and the Crab Pulsar.
The first seven columns give the source name, the observed bands, the
observed maximum delay $\Delta t$ between the rise of the light curve
at two different frequencies, the reference for the observation, the low
frequency, the high frequency, and the source distance D.  Cosmological
distances were calculated from the look-back time as a function of red
shift for the most conservative reasonable case of $H_{0} = 80 km \cdot 
s^{-1} Mpc^{-1}$ and $\Omega = 1$.  Observed GRBs are known to have a
relatively narrow luminosity function (as required by the fast turnover in
the LogN-LogP relation \cite{horack}), limits on host galaxies forces the
peak luminosity to be much greater than $10^{58} photons \cdot
s^{-1}$\cite{schaefer2}, and the three bursts with measured red shifts all
have luminosities $> 4 \times 10^{57} photons \cdot s^{-1}$
\cite{schaefer2}. So I have adopted
a peak luminosity of $10^{57} photons \cdot s^{-1}$ as a conservative 
minimum for use in Fenimore's peak flux versus red shift relation
\cite{fenimore}.

All the events at cosmological distances are explosive, so presumably the
light curves of different wavelengths will start to rise simultaneously in
the absence of dispersion.  The lack of flux or the lack of observations
might produce an apparent delay not associated with dispersion.  Also, the
explosive system might have an intrinsic delay in emission which could
counter any effects of dispersion.  Nevertheless, I will assume that no
conspiracies of delays hides the effect of dispersion.  The number of
strict limits from widely disparate classes of events argues that any such
conspiracy is unlikely.

The frequency dependence of `c' is not known, so no single number can
represent the limit on variations in the speed of light.  The most model
independent parameter is one with no reference to the observed
frequencies, $\Delta c/c$.  This limit on $\Delta c/c$ will be $c
\cdot \Delta t/D$. For a general dispersion relation like 
$V = c(1+A\nu^{-2})^{0.5}$ with V as the velocity of light with frequency
$\nu$ and both c and A as constants \cite{jackson} \cite{feinberg}, the
limit on A is $(2\cdot c \cdot \Delta t/D)\cdot (\nu_{2}^{-2} -
\nu_{1}^{-2})^{-1}$ for observations at frequencies $\nu_{1}$ and
$\nu_{2}$. This dispersion can be related to the photon mass as
$m_{\gamma} = A^{0.5}h/c^{2}$.  For quantum gravity models \cite{camelia},
the characteristic energy $E_{QG}$ is greater than $h \cdot (\nu_{2} -
\nu_{1}) \cdot D/(c \cdot \Delta t)$. The limits on $\Delta c/c$,
$m_{\gamma}$, and $E_{QG}$ for each event are presented in columns 
8-10 of Table 1.  

The limits in Table 1 are many orders of magnitude past those from the
Crab Pulsar.  The strictest limit on $\Delta c/c$ is $6.3 \times 10^{-21}$
for GRB 930229, with second place at $1.4 \times 10^{-20}$ for GRB 910711.
The two lowest limits on the photon mass are $4.2 \times 10^{-44} g$ and
$1.5 \times 10^{-43} g$ for GRB 980703 and GRB 970508, both from radio to
gamma ray constraints for bursts with measured red shifts.  The tightest
limit on $E_{QG}$ is $8.3 \times 10^{16}$ GeV for GRB 930131 followed
closely by $6.0 \times 10^{16}$ GeV for Mkn 421.

The new limit on $\Delta c/c$ is close to $10^{4}$ times better than the
Crab Pulsar limit, and it is comforting to know that Einstein's postulate
is vindicated to this level. The limit on the photon mass is $10^{4}$
times worse
than that obtained by considering the Earth's magnetic field
\cite{goldhaber}, however
this result uses virtual photons with $\nu = 0$ so an independent method
is still of interest \cite{feinberg}.  The limits on $E_{QG}$ are not yet
close to
the expected $\sim 10^{19}$ GeV \cite{camelia} \cite{biller}. 

Significant improvements in limits on $\Delta c/c$ are unlikely since it
we are already dealing with delays of under a millisecond out of a Hubble
time.  Limits on the photon mass can be improved by several orders of
magnitude with radio detection of prompt emission from GRBs (as with the
FLIRT telescope \cite{balsano} ) or with radio studies of millisecond
pulsars. The detection of $\sim 100$ GeV photons near the start of a GRB
of moderate brightness is possible with the GLAST satellite mission
\cite{bonnell}, and this would test the expected quantum gravity
threshold.

\clearpage

\begin{table}
\caption{Limits on $\Delta c/c$, $m_{\gamma}$, and $E_{QG}$ from explosive 
events at high redshift.}
\begin{center}
\begin{tabular}{cccccccccc}
EVENT & Bands & $\Delta t$ & Ref. & $\nu_{1}(Hz)$ & $\nu_{2}(Hz)$ &
D (Mpc) & $\Delta c/c$ & $m_{\gamma}$(g) & $E_{QG}$(GeV) \\
\tableline
GRB 930229 & $\gamma - \gamma$ & 0.5 ms & \cite{schaefer} & 
$7.2 \times 10^{18}$ & $4.8 \times 10^{19}$ & 791 & 
$6.3 \times 10^{-21}$ & $6.1 \times 10^{-39}$ & $2.7 \times 10^{16}$ \\
GRB 910711 & $\gamma - \gamma$ & 2 ms & \cite{bhat} &
$7.2 \times 10^{18}$ & $1.2 \times 10^{20}$ & 1413 &
$1.4 \times 10^{-20}$ & $9.0 \times 10^{-39}$ & $3.3 \times 10^{16}$ \\
GRB 910625 & $\gamma - \gamma$ & 4 ms & $\cdots$ &
$7.2 \times 10^{18}$ & $1.2 \times 10^{20}$ & 1344 &
$3.0 \times 10^{-20}$ & $1.3 \times 10^{-38}$ & $1.6 \times 10^{16}$ \\
GRB 910607 & $\gamma - \gamma$ & 8 ms & $\cdots$ &
$7.2 \times 10^{18}$ & $1.2 \times 10^{20}$ & 1677 &
$4.8 \times 10^{-20}$ & $1.7 \times 10^{-38}$ & $9.9 \times 10^{15}$ \\
GRB 930131 & $\gamma - GeV$ & 25 ms & \cite{sommer} &
$7.2 \times 10^{18}$ & $1.9 \times 10^{22}$ & 260 &
$9.6 \times 10^{-19}$ & $7.4 \times 10^{-38}$ & $8.3 \times 10^{16}$ \\
GRB 930131 & $\gamma - GeV$ & 0.5 s & \cite{sommer} &
$7.2 \times 10^{18}$ & $1.1 \times 10^{23}$ & 260 &
$1.9 \times 10^{-17}$ & $3.3 \times 10^{-37}$ & $2.4 \times 10^{16}$ \\
GRB 940217 & $\gamma - GeV$ & 4800 s & \cite{hurley} &
$7.2 \times 10^{18}$ & $4.3 \times 10^{24}$ & 385 &
$1.2 \times 10^{-13}$ & $2.7 \times 10^{-35}$ & $1.4 \times 10^{14}$ \\
GRB 970508 & $X - \gamma$ & 5.6 hrs & \cite{piro} &
$2.4 \times 10^{17}$ & $1.2 \times 10^{20}$ & 1493 &
$1.4 \times 10^{-13}$ & $9.2 \times 10^{-37}$ & $3.7 \times 10^{9}$ \\
GRB 970508 & $U - \gamma$ & 4.4 hrs & \cite{castro} &
$8.2 \times 10^{14}$ & $1.2 \times 10^{20}$ & 1493 &
$1.1 \times 10^{-13}$ & $2.8 \times 10^{-39}$ & $4.7 \times 10^{9}$ \\
GRB 970508 & $Radio - \gamma$ & 121 hrs & \cite{frail} &
$8.6 \times 10^{9}$ & $1.2 \times 10^{20}$ & 1493 &
$2.9 \times 10^{-12}$ & $1.5 \times 10^{-43}$ & $1.7 \times 10^{8}$ \\
GRB 980703 & $X - \gamma$ & 22 hrs & \cite{galama} &
$2.4 \times 10^{17}$ & $1.2 \times 10^{20}$ & 1592 &
$5.0 \times 10^{-13}$ & $1.8 \times 10^{-36}$ & $1.0 \times 10^{9}$ \\
GRB 980703 & $I - \gamma$ & 21 hrs & \cite{vreeswijk} &
$3.3 \times 10^{14}$ & $1.2 \times 10^{20}$ & 1592 &
$4.8 \times 10^{-13}$ & $2.4 \times 10^{-39}$ & $1.0 \times 10^{9}$ \\
GRB 980703 & $Radio - \gamma$ & 29 hrs & \cite{frail2} &
$5.0 \times 10^{9}$ & $1.2 \times 10^{20}$ & 1592 &
$6.6 \times 10^{-13}$ & $4.2 \times 10^{-44}$ & $7.6 \times 10^{8}$ \\                                                                     
SN 1997ap & $I-R$ & 240 hrs & \cite{perlmutter} &
$3.3 \times 10^{14}$ & $4.3 \times 10^{14}$ & 1489 &
$5.8 \times 10^{-12}$ & $1.3 \times 10^{-38}$ & $6.8 \times 10^{1}$ \\ 
SN 1994G & $I-R$ & 72 hrs & \cite{perlmutter2} &
$3.3 \times 10^{14}$ & $4.3 \times 10^{14}$ & 1029 &
$2.5 \times 10^{-12}$ & $8.8 \times 10^{-28}$ & $1.6 \times 10^{2}$ \\
Mkn 421 & $TeV-TeV$ & 280s & \cite{biller} &
$1.2 \times 10^{26}$ & $4.8 \times 10^{26}$ & 112 &
$2.5 \times 10^{-14}$ & $2.1 \times 10^{-28}$ & $6.0 \times 10^{16}$ \\
Crab & $V-U$ & 0.01ms & \cite{warner} &
$5.5 \times 10^{14}$ & $8.2 \times 10^{14}$ & 0.002 &
$5.0 \times 10^{-17}$ & $5.4 \times 10^{-41}$ & $2.3 \times 10^{7}$ \\
\end{tabular}
\end{center}
\end{table}


\begin{references}

\bibitem{fukuda} Y. Fukuda {\em et al.}, Phys. Rev. Lett. {\bf 81}, 1562
(1998).

\bibitem{camelia} G. Amelino-Camelia {\em et al.}, Nature {\bf 393}, 763
(1998).

\bibitem{gambini} R. Gambini and J. Pullin, gr-qc/9809038 (1998).

\bibitem{jackson} A. P. French, {\em Special Relativity} (W. W. Norton,
New York, 1968); J. D. Jackson, {\em Classical Electrodynamics} (Wiley,
New York, 1975).

\bibitem{hall} J. L. Hall, in {\em Atomic Masses and Fundamental
Constants}, edited by J. H. Sanders and A. H. Wapstra (Plenum, New York,
1976) p. 322; R. H. Garstang, A. J. {\bf 79}, 1260 (1974).

\bibitem{lovell} B. Lovell, F. Whipple, and L. Solomon, Nature {\bf 202},
377 (1964).

\bibitem{warner} B. Warner and R. Nather, Nature {\bf 222}, 157 (1969).

\bibitem{thompson} D. J. Thompson, {\em in Pulsars, Problems and
Progress}, IAU Coll. 160, edited by S. Johnston, M. Walker, and M. Bailes
(San Francisco, ASP, 1996), p. 307.

\bibitem{biller} S. D. Biller {\em et al.}, Phys. Rev. Lett. (submitted,
gr-qc/9810044).

\bibitem{schaefer} B. E. Schaefer and K. C. Walker, Ap. J. Lett.
(submitted, astro-ph/9810270).

\bibitem{horack} J. M. Horack, A. G. Emslie, and C. A. Meegan, Ap. J.
{\bf 426}, L5 (1994).

\bibitem{schaefer2}  B. E. Schaefer, Ap. J. Lett. (submitted,
astro-ph/9810424).

\bibitem{fenimore} E. E. Fenimore et al., Nature {\bf 357}, 140 (1992).

\bibitem{feinberg} G. Feinberg, Science {\bf 166}, 879 (1969); Z. Bay, and
J. A. White, Phys. Rev. D {\bf 5}, 796 (1972).

\bibitem{goldhaber} A. S. Goldhaber and M. M. Nieto, Rev. Mod. Phys.
{\bf 43}, 277 (1971).

\bibitem{balsano} R. J. Balsano et al., in {\em Gamma Ray Bursts}, AIP
Conf. Proc. {\bf 428}, edited by C. A. Meegan, R. D. Preece, and T. M.
Koshut (AIP, New York, 1998), p. 585.

\bibitem{bonnell} J. T. Bonnell et al., in {\em Gamma Ray Bursts}, AIP
Conf. Proc. {\bf 428}, edited by C. A. Meegan, R. D. Preece, and T. M.
Koshut (AIP, New York, 1998), p. 884.

\bibitem{bhat} P. N. Bhat {\em et al.}, Nature {\bf 359}, 217 (1992).

\bibitem{sommer} M. Sommer {\em et al.}, Ap. J. {\bf 422}, L63 (1994).

\bibitem{hurley} K. Hurley {\em et al.}, Nature {\bf 372}, 652 (1994).

\bibitem{piro} L. Piro {\em et al.}, IAU Circ. 6656 (1997).

\bibitem{castro} A. J. Castro-Tirado {\em et al.}, IAU Circ. 6657 (1997).

\bibitem{frail} D. A. Frail {\em et al.}, Nature 389, 261 (1997).

\bibitem{galama} T. J. Galama {\em et al.}, GCN\#127 (1998).

\bibitem{vreeswijk} P. M. Vreeswijk {\em et al.}, GCN\#132 (1998).

\bibitem{frail2} D. A. Frail {\em et al.}, GCN \#128 (1998).

\bibitem{perlmutter} S. Perlmutter {\em et al.}, Nature {\bf 391}, 51
(1998).

\bibitem{perlmutter2} S. Perlmutter {\em et al.}, Ap. J. {\bf 483}, 565
(1997).

\end{references}
\end{document}